\documentclass[epj]{svjour}
\usepackage{graphics}
\newcommand{\beq}{\begin{equation}}
\newcommand{\eeq}{\end{equation}}
\begin{document}
\title{Theoretical Analysis of Two-Color Ghost Interference}
\author{Sheeba Shafaq\inst{1} \and Tabish Qureshi\inst{2}}

\institute{Department of Physics, Jamia Millia Islamia, New Delhi,
\email{shafaqsheeba1@gmail.com}
\and 
Centre for Theoretical Physics, Jamia Millia Islamia, New Delhi,
\email{tabish@ctp-jamia.res.in}}

\abstract{
Recently demonstrated ghost interference using correlated photons of different
frequencies, has been theoretically analyzed. The calculation predicts an
interesting nonlocal effect: the fringe width of the ghost interference 
depends not only on the wave-length of the photon involved, but also on the
wavelength of the other photon with which it is entangled. This feature,
arising because of different frequencies of the entangled photons, was
hidden in the original ghost interference experiment. This prediction can be
experimentally tested in a slightly modified version of the experiment.}


\maketitle

\section{Introduction}

\begin{figure}
\centerline{\resizebox{8.0cm}{!}{\includegraphics{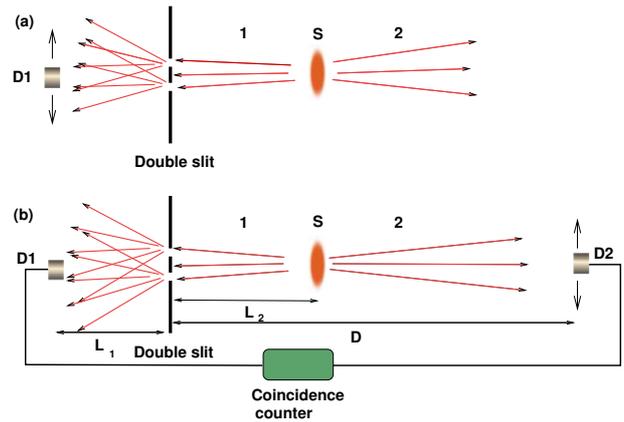}}}
\caption{Ghost interference: An SPDC source generates photon pairs - one goes left,
and the other right. (a) Putting a double slit in the path of photon 1 results
in no interference. (b)  Detecting photon 2 in coincidence with a
{\em fixed} detector D1 clicking, results in a ghost interference \cite{ghost}.}
\label{ghostfig}
\end{figure}

The nonlocal nature of quantum correlations that exist in spatially separated
entangled particles, has been a subject of attention since the time it was
first pointed out by Einstein, Podolsky and Rosen \cite{epr}. The most
dramatic experimental demonstration of these correlations has been provided
by the {\em ghost interference} experiment by Strekalov et al.\cite{ghost}.
In this experiment, 
a Spontaneous Parametric Down-Conversion (SPDC) source S sends out pairs of
entangled photons, which we call photon 1 and photon 2 (see Fig. \ref{ghostfig}).
A double-slit is placed in the path of photon 1. The most intriguing part of
the experiment is that photon 2, which doesn't have any double-slit in
its path, shows an interference when detected in coincidence with a fixed
detector D1 for photon 1 (see Fig. \ref{ghostfig}). This phenomenon was appropriately
called ghost interference, and has attracted considerable research
attention \cite{ghostimaging,rubin,zhai,jie,zeil2,pravatq}.

Of late, people have explored another way of generating photon pairs,
based on spontaneous four-wave mixing (SFWM) in an atomic ensemble
\cite{sfwm1,sfwm2,sfwm3}. The photons generated by this technique have very
narrow bandwidth. Recently a ghost interference experiment has been carried out
using photon pairs generated via SFWM \cite{twocolor}. The novel feature of
this experiment is that the correlated photons in a pair are of different
color, with wavelengths $\lambda_1=1530$ nm and $\lambda_2=780$ nm, and
has been called {\em two-color ghost interference} by the authors. This
ghost interference observed with photons with $\lambda_2=780$ nm was 
explained by the authors using a simple geometrical analysis \cite{twocolor},
along the lines of the geometrical analysis in the original ghost interference
experiment \cite{ghost}.

It has been shown earlier that a more thorough analysis, which goes beyond the
simple geometrical argument, is needed to fully understand the phenomenon of
ghost interference \cite{pravatq}. Here we carry out a wave-packet analysis for
the two-color ghost interference experiment for a more quantitative 
understanding of the phenomenon. This analysis unveils an interesting
effect which can be
tested in a modified version of the two-color ghost interference.

\section{Ghost interference}

We begin by describing the ghost interference experiment of Strekalov et al.
\cite{ghost}. Entangled photons, 1 and 2, are emitted from a source S.
Photon 1 passes through a double-slit to reach detector D1, whereas photon 2
travels to detector D2 unhindered.  The results of the experiment are as
follows.

\noindent ({\it a}) When photons 1 are detected 
using a detector placed behind the double-slit, no first order interference 
is observed for photon 1. This is surprising because interference is
generally expected when photons pass through a double-slit. 
For photons 2, first order interference is neither expected, nor observed.

\noindent ({\it b}) When photons 2 are detected {\em in coincidence
with a fixed detector behind the double slit registering photon 1}, an
interference
pattern which is very similar to a double-slit interference pattern is 
observed,  
even though there is no double-slit in the path of photon 2. Changing the
position of the fixed detector does not change the interference pattern, but
only shifts it. 

Another curious feature is that the fringe-width of the interference pattern
is the same as what one would
observe if one were to replace the photon detector D1 behind the 
double slit, by a source of light, and the SPDC source were absent. In 
other words, 
the standard Young's double slit interference formula works, if the distance
is taken to be the distance between the screen (detector) on which photon 2
registers, right through the SPDC source crystal, to the double slit. Photon
2 never passes through the region between the source S and the double 
slit. 

Ghost interference is now well understood as a combined effect
of a virtual double-slit formation for photon 2, due to entanglement, and the
{\em quantum erasure} of which-path information by the fixed detector
\cite{pravatq}. The non-observation of the first order interference for
photon 1 is due to the which-path information carried by photon 2 which, in
principle, can be used to tell which slit photon 1 passed
through \cite{pravatq}. By Bohr's principle of complementarity, no
interference can be observed in such a situation.

There is a closely related phenomenon called ghost imaging which was first
unraveled using entangled photons \cite{imaging}. Here, two photons from
a source go in different directions. On one side, the photons are made to
pass through a filter and all those that pass through the filter are collected
by a ``bucket" detector. On the other side, the photons do not go through any
filter, but are detected by a point detector which scans the direction
perpendicular to the motion of photons. This detector, counted in coincidence
with the other detector, reproduces the image of the filter which is 
present in the path of the other detector. As this photon does not pass through
any filter, the pattern reproduced is named ghost image.

Later on it was
demonstrated, theoretically and experimentally, that ghost imaging can also be
done using pseudothermal light
\cite{boyd1,boyd2,lugiato1,lugiato2,lugiato3,shih1,lugiato4,cheng,bai,basano}.
Ghost imaging with entangled photons of different wavelengths has also been
studied \cite{2colghost1,2colghost2,2colghost3}.
Whether quantum correlations are needed to explain ghost imaging or are classical
correlations in light sufficient, is a topic which has been hotly debated
\cite{shih2,shapiro1,shih3,shapiro2}. The present consensus is that
pseudothermal ghost imaging can be understood as arising from correlations
between classical speckle patterns formed on the object and the reference
detector, or from two-photon interference between the photons falling on
the two detectors. Here quantum theory is not really necessary. For
describing ghost imaging from SPDC photons quantitatively, quantum theory
is necessary.

For some time it was believed that although ghost imaging is possible with
pseudothermal light, ghost interference with pseudothermal light may not
be possible. However, later on it was shown, theoretically and experimentally,
that ghost interference is also possible with pseudothermal
light \cite{lugiato4,cheng,bai,basano}.
Although the ghost interference demonstrated using pseudothermal
light looks qualitatively similar to that obtained with entangled photons,
the origins of the two are different. The ghost interference as demonstrated in
\cite{ghost} needs a quantum explanation, especially in the light of the
curious distance appearing in the Young's double-slit formula that works in
that experiment.

\begin{figure}
\centerline{\resizebox{8.0cm}{!}{\includegraphics{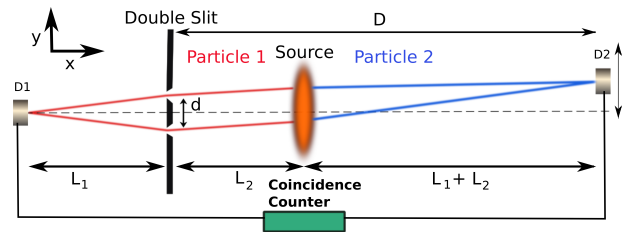}}}
\caption{ Two-color ghost interference: A source generates pairs of photons of
different wavelengths - one goes left, and the other right. Counting of
photon 2 in coincidence with a {\em fixed} detector D1 clicking, results
in a ghost interference \cite{twocolor}.}
\label{twocolfig}
\end{figure}

\section{Theoretical analysis}

The two-color ghost interference experiment can be schematically represented
as shown in Fig. \ref{twocolfig}. Ding et al. have used a converging lens before the 
detector D2 \cite{twocolor}, which was not there in the original ghost
interference experiment \cite{ghost}. We will carry out our analysis without
the converging lens. The use of photons is not essential to ghost
interference. Any two entangled particles should lead to the same
phenomenon. There is an experiment with electrons which shows similar
quantum correlations \cite{neder}. In addition, there have been proposals
to observe ghost interference with entangled massive particles \cite{zeil2}.
We will thus carry out our analysis using two
entangled massive particles. The two photons having different frequencies
would translate to the two particles having different masses. As we will be
using wave-packets in our analysis, the connection to photons can be made
easily. 

We assume that the two detectors D1 and D2 sit at equal distance from the
source which produces pairs of particles, which we label 1 and 2, with
mass $m_1$ and $m_2$, which move along --ve and +ve x-axis, respectively.
The source is assumed to produce particles in the following initial state:
\begin{equation}
\Psi(y_1,y_2) = C\!\int_{-\infty}^\infty dp
e^{-p^2/4\sigma^2}e^{-ipy_2/\hbar} e^{i py_1/\hbar}
e^{-{(y_1+y_2)^2\over 4\Omega^2}}, \label{state}
\end{equation}
where $C$ is a normalization constant. The
$e^{-(y_1+y_2)^2/4\Omega^2}$ term is required so that the state (\ref{state}) 
is normalized in $y_1$ and $y_2$. 
This is a generalized form of the EPR state \cite{epr}, and reduces to it
in the limit $\sigma\to\infty,~\Omega\to\infty$. 
As is obvious from the form of the state, we are only interested in the
entanglement in the $y$ degrees of freedom of the particles. The particles
are assumed to travel along x-axis with constant velocities. This motion
is not interesting, as far as entanglement is concerned, and will be 
ignored in the calculation. We assume that we know the x-positions of the
particles after any given time, from classical motion.

Integration over $p$ can be performed to obtain:
\begin{equation}
\Psi(y_1,y_2) = \sqrt{ {\sigma\over \pi\hbar\Omega}}
 e^{-(y_1-y_2)^2\sigma^2/\hbar^2}
e^{-(y_1+y_2)^2/4\Omega^2} .
\label{newstate}
\end{equation}
The physical meaning of the constants $\sigma$ and $\Omega$ will become clear
if we calculate the uncertainty in position and momentum of the two particles.
The uncertainty in momenta of the two particles is given by
\begin{equation}
\Delta p_{1y} = \Delta p_{2y} = 
{1\over 2}\sqrt{\sigma^2 + {\hbar^2\over 4\Omega^2}}. \label{dp}
\end{equation}
The position uncertainty of the two particles is given by
\begin{equation}
\Delta y_1 = \Delta y_2 = \sqrt{\Omega^2+\hbar^2/4\sigma^2}.
\end{equation}
With time, the particles travel along the positive and
negative x-axis. The motion in the x-direction is disjoint from the
evolution in the y-direction, and is unaffected by entanglement. 

The state evolves for a time $t_0$ before particle 1 reaches the
double-slit.
The state of the entangled system, after this time evolution, can be
calculated using the Hamiltonian governing the time evolution, given by
\begin{equation}
\hat{H} = -{\hbar^2\over 2m_1} {\partial^2\over \partial y_1^2} 
            -{\hbar^2\over 2m_2} {\partial^2\over \partial y_2^2} \label{H}
\end{equation}
After a time $t_0$, (\ref{newstate}) assumes the form
\begin{eqnarray}
\Psi(y_1,y_2,t_0) &=& {1\over \sqrt{{\pi}(\Omega+{i\hbar t_0\over 2M\Omega})
(\hbar/\sigma + {2i\hbar t_0\over \mu\hbar/\sigma})}}\nonumber\\
&&\exp\left[{-(y_1-y_2)^2\over \hbar^2/\sigma^2  + {2i\hbar t_0\over \mu}}
\right]
\exp\left[{-(y_1+y_2)^2\over \left(4\Omega^2  +
{2i\hbar t_0\over M}\right)} \right],\nonumber\\
\label{Psit}
\end{eqnarray}
where $M=m_1+m_2$ and  $\mu = {m_1m_2\over m_1+m_2}$.
We wish to point out that the use of (\ref{H}) is not an absolute necessity
for obtaining  the time evolution of the state. For example,  if one
considers the particle to be an envelope of waves , the time evolution can be
obtained easily. In that case, $\left({d^2\omega(k)\over dk^2}\right)_{k_0}$,
where $k_0$ is the wave-vector value where $\omega(k)$ peaks, plays the role
of $\hbar/m$.
The time evolution for a photon state can be obtained similarly \cite{mandel}.

In order to incorporate the effect of the double-slit on the system of
entangled particles, we assume that the double-slit allows portions of the
wavefunction in front of the slit to pass through, and blocks the other
portions. Let us assume that what
emerges from a single slit is a localized Gaussian packet, whose width is the 
width of the slit. So, if the two slits are A and B, the packets which
pass through will be, say, $|\phi_A(y_1)\rangle$ and $|\phi_B(y_1)\rangle$,
respectively. The portion of particle 1 which gets blocked is, say,
$\chi(y_1)$. As these three states are obviously orthogonal, any state of
particle 1 can be expanded in terms of these. We can thus write:
\begin{equation}
|\Psi(y_1,y_2,t_0)\rangle = |\phi_A\rangle\langle\phi_A|\Psi\rangle
+ |\phi_B\rangle\langle\phi_B|\Psi\rangle +
|\chi\rangle\langle\chi|\Psi\rangle . \label{slit}
\end{equation}
The corresponding states of particle 2 can be calculated as
\begin{eqnarray}
\psi_A(y_2) &=& \langle\phi_A(y_1)|\Psi(y_1,y_2,t_0)\rangle \nonumber\\
\psi_B(y_2) &=& \langle\phi_B(y_1)|\Psi(y_1,y_2,t_0)\rangle \nonumber\\
\psi_\chi(y_2) &=& \langle\chi(y_1)|\Psi(y_1,y_2,t_0)\rangle \label{psi}
\end{eqnarray}

So, the state we get after particle 1 crosses the double-slit is:
\begin{equation}
|\Psi(y_1,y_2)\rangle = |\phi_A\rangle|\psi_A\rangle
+ |\phi_B\rangle|\psi_B\rangle +
|\chi\rangle|\Psi_\chi\rangle ,
\end{equation}
where $|\phi_A\rangle$ and $|\phi_B\rangle$ are states of particle 1,
and $|\psi_A\rangle$ and $|\psi_B\rangle$ are states of particle 2.
The first two terms represent the amplitudes of particle 1 passing through
the slits, and the last term represents the amplitude of it getting
reflected/blocked.
Because of the linearity of Schr\"odinger equation, these two parts of
the wavefunction will evolve independently, without affecting each
other. Since we are only interested in situations where particle 1 passes
through the slit, we might as well throw away the term which represents
particle 1 not passing through the slits. If we do that, we have to renormalize
the remaining part of the wavefunction, which looks like
\begin{equation}
|\Psi(y_1,y_2)\rangle = {1\over C} (|\phi_A\rangle|\psi_A\rangle
+ |\phi_B\rangle|\psi_B\rangle),
\end{equation}
where $C = \sqrt{\langle\psi_A|\psi_A\rangle + \langle\psi_B|\psi_B\rangle}$.

In the following, we assume that $|\phi_A\rangle$, $|\phi_B\rangle$, are
Gaussian wave-packets:
\begin{eqnarray}
\phi_A(y_1) &=& {1\over(\pi/2)^{1/4}\sqrt{\epsilon}} e^{-(y_1-y_0)^2/\epsilon^2}
,\nonumber\\
\phi_B(y_1) &=& {1\over(\pi/2)^{1/4}\sqrt{\epsilon}} e^{-(y_1+y_0)^2/\epsilon^2},
\end{eqnarray}
where $\pm y_0$ is the y-position of slit A and B, respectively, and $\epsilon$
their widths. Thus, the distance between the two slits is $2 y_0 \equiv d$.

Using (\ref{psi}) and (\ref{Psit}), wavefunctions $|\psi_A\rangle$,
$|\psi_B\rangle$ can be calculated, which, after normalization, have the form
\begin{equation}
\psi_A(y_2) = C_2 e^{-{(y_2 - y_0')^2 \over \Gamma}},~~~
\psi_B(y_2) = C_2 e^{-{(y_2 + y_0')^2 \over \Gamma}} ,
\end{equation}
where $C_2 = (2/\pi)^{1/4}(\sqrt{\Gamma_R} + {i\Gamma_I\over\sqrt{\Gamma_R}})^{-1/2}$,
\begin{equation}
y_0' = {y_0 \over {4\Omega^2\sigma^2/\hbar^2+1\over
4\Omega^2\sigma^2/\hbar^2-1} + {4\epsilon^2 \over 4\Omega^2-\hbar^2/\sigma^2}},
\end{equation}
and
\begin{equation}
\Gamma = \frac{{\hbar^2\over\sigma^2}(1+{\epsilon^2+2i\hbar t_0/M\over 4\Omega^2})
 + \epsilon^2+2i\hbar t_0/\mu +{2i\hbar t_0\over 4\Omega^2}({1\over M}+
{1\over\mu})}{1 + {\epsilon^2\over\Omega^2}+{i\hbar t_0\over 2\Omega^2}({1\over M}+{1\over\mu}) +
{\hbar^2\over 4\Omega^2\sigma^2}}
\end{equation}

Thus, the state which emerges from the double slit, has the following form
\begin{eqnarray}
\Psi(y_1,y_2) &=& c~e^{-(y_1-y_0)^2/\epsilon^2}
e^{-{(y_2 - y_0')^2 \over \Gamma}}\nonumber\\
&&+ c e^{-(y_1+y_0)^2/\epsilon^2}
e^{-{(y_2 + y_0')^2 \over \Gamma}} \label{virtual},
\end{eqnarray}
where $c = 
(1/\sqrt{\pi})(\sqrt{\Gamma_r} + {i\Gamma_i\over\sqrt{\Gamma_r}})^{-1/2}$.
Equation (\ref{virtual}) represents two wave-packets of particle 1,
of width $\epsilon$, and localized at $\pm y_0$, entangled with two
wave-packets of particle 2, of width
${\sqrt{2}|\Gamma|\over\sqrt{\Gamma+\Gamma^*}}$, localized at
$\pm y_0'$.

If the two wave-packets of particle 2 are orthogonal, 
the amplitudes of particle 1 passing through the two slits are correlated with
two distinguishable states of particle 2. In principle, one can make a
measurement on {\em particle 2} and find out which slit {\em particle 1}
passed through. Bohr's Complementarity principle says that no interference
can be observed in such a situation. So, no first order interference can be
seen in particle 1 because particle 2 carries the ``which-way" information
about particle 1. This is the real reason for photon 1 not showing
first order interference in the original ghost interference experiment
\cite{ghost} and also in the two-color ghost interference
experiment \cite{twocolor}.

\subsection{Entanglement and virtual double-slit}

From (\ref{virtual}) one can see that the state of particle 2 also involves
two spatially separated, localized Gaussians, correlated with states of
particle 1. So, because of entanglement, particle 2 also behaves as if it
has passed through a double-slit of separation $2y_0'$. In other words,
because of entanglement, particle 1 passing through the double-slit,
creates a {\em virtual double-slit} for particle 2. This view is in agreement
with the observed optical imaging by means of entangled photons \cite{imaging}.
It appears natural that
particle 2, passing through this virtual double-slit, should show an
interference pattern. However, this can happen only when the wave-packets
overlap, after evolving in time.

Before reaching detector D2, particle 2 evolves for a time $t$.
The time evolution, governed by (\ref{H}), transforms the state (\ref{virtual})
to
\begin{eqnarray}
\Psi(y_1,y_2,t) &=& 
C_t \exp\left[{{-(y_1-y_0)^2\over\epsilon^2+2i\hbar t/m_1}}\right]
 \exp\left[{{-(y_2 - y_0')^2 \over \Gamma+2i\hbar t/m_2}}\right]
\nonumber\\
&&+ C_t \exp\left[{{-(y_1+y_0)^2\over\epsilon^2+2i\hbar t/m_1}}\right]
 \exp\left[{{-(y_2 + y_0')^2 \over \Gamma+2i\hbar t/m_2}}\right],\nonumber\\
\end{eqnarray}
where 
\begin{equation}
C(t) = {1\over \sqrt{\pi}\sqrt{\epsilon+2i\hbar t/m_1\epsilon}
\sqrt{\sqrt{\Gamma_r}+(\Gamma_i+2i\hbar t/m_2)/\sqrt{\Gamma_r}}}.
\end{equation}
If the correlation between the particles is good, one can make the
following approximation:
$\Omega \gg \epsilon$ and $\Omega \gg \hbar/\sigma$. In this
limit,
\begin{equation}
\Gamma \approx \gamma^2 + 2i\hbar t_0/\mu,~~~ y_0' \approx y_0,
\end{equation}
where $\gamma^2 = \epsilon^2 + \hbar^2/\sigma^2$.
We are now in a position to calculate the probability of finding particle 1
at $y_1$ and particle 2 at $y_2$. 

Before we do that, it will be useful to translate our results to the language
of the optical experiment where one talks of wavelength and measures distances.
If a particle of mass $m$ travels along the x-axis, with a momentum $p$,
in time $t$, it travels a distance (say) $L$. So, we can write
$\hbar t/m = \hbar vt/p = \lambda vt/2\pi = \lambda L/2\pi$, where $v$
is the velocity of the particle, and $\lambda$ its d'Broglie wavelength.

Using this strategy, for photons we can write
$\hbar t_0/m_1 = \lambda_1 L_2/2\pi$,
$\hbar t_0/m_2 = \lambda_2 L_2/2\pi$,
$\hbar t/m_1 = \lambda_1 L_1/2\pi$,
$\hbar t/m_2 = \lambda_2 L_1/2\pi$.
Also, for compactness we will use the following notation,
\begin{equation}
D \equiv L_1+2L_2,~~~~ L \equiv L_1+L_2,~~~~~~ d \equiv 2y_0.
\end{equation}

The probability density of finding particle 1 at $y_1$ and particle 2 at $y_2$
is given by
%\begin{widetext}
\begin{eqnarray}
P(y_1,y_2) &=& |\Psi(y_1,y_2,t)|^2 \nonumber\\
&=& |C_t|^2 \left(  
\exp\left[-{2(y_1-y_0)^2\over\epsilon^2+{\lambda_1^2 L_1^2\over\pi^2\epsilon^2}}
-{2(y_2 - y_0)^2 \over \gamma^2+{(\lambda_2 L+\lambda_1 L_2)^2\over\pi^2\gamma^2}}\right]\right. \nonumber\\
&&+ \exp\left[-{2(y_1+y_0)^2\over\epsilon^2+({\lambda_1 L_1\over\pi\epsilon})^2}
-{2(y_2 + y_0)^2 \over \gamma^2+({\lambda_2 L+\lambda_1 L_2\over\pi\gamma})^2}\right]
\nonumber\\
&&+ \exp\left[-{2(y_1^2+y_0^2)\over\epsilon^2+({\lambda_1 L_1\over\pi\epsilon})^2}
-{2(y_2^2 + y_0^2) \over \gamma^2+({\lambda_2 L+\lambda_1 L_2\over\pi\gamma})^2}\right]\nonumber\\
&&\left.\times 2\cos\left[\theta_1 y_1 + \theta_2 y_2\right]\right),
\label{pattern}
\end{eqnarray}
%\end{widetext}
where 
\begin{equation}
\theta_1 = {2d\lambda_1 L_1/\pi\over \epsilon^4 + \lambda_1^2L_1^2/\pi^2},~~~
\theta_2 = {2\pi d[\lambda_2 L+\lambda_1L_2)]\over \gamma^4\pi^2 +
(\lambda_2 L+\lambda_1L_2)^2},
\end{equation}
and
\begin{equation}
C_t = {1\over \sqrt{\pi}\sqrt{\epsilon+{i\lambda_1L_1\over\pi\epsilon}}
\sqrt{\gamma+{i\lambda_2 L+i\lambda_1L_2\over\pi\gamma}}}.
\end{equation}

\begin{figure}[h]
\centerline{\resizebox{9.0cm}{!}{\includegraphics{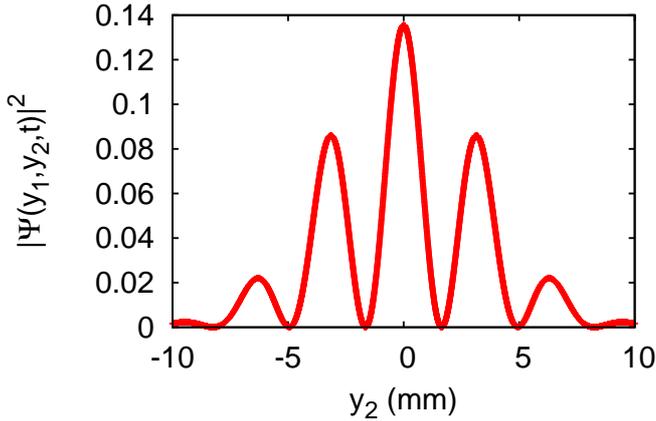}}}
\caption{ Probability density of particle 2 as a function of the position of
detector D2, with D1 fixed at $y_1=0$, for $\lambda_1=1530$ nm,
$\lambda_2=780$ nm, $D=1.8$ m,
$L_1=1.15$ m, $L_2 = 32.5$ cm, $d=0.5$ mm, $\epsilon=0.1$ mm and
$\gamma=0.11$ mm }
\label{graph}
\end{figure}

In the two-color ghost interference experiment, detector D1 is kept fixed
and detector D2 is scanned along the y-axis. If we fix $y_1$, the cosine
term in (\ref{pattern}) represents oscillations as a function of $y_2$. 
This means that if particle 2 is detected in coincidence with a fixed D1,
it will show an interference. The exponential terms represent overall
Gaussian envelope on the interference pattern. The expression (\ref{pattern})
should correctly describe the two-color ghost interference.
The probability density given by (\ref{pattern}), when plotted against 
the position of D2, yields an interference pattern (see Fig. \ref{graph}).
The fringe width of the pattern for particle 2 is given by
\begin{equation}
w_2 = {2\pi \over \theta_2} = {\lambda_2 D\over d} + {(\lambda_1-\lambda_2)L_2\over d}+
{\gamma^4\pi^2\over d\lambda_2 D + d(\lambda_1-\lambda_2)L_2}
\end{equation}
For $\pi\gamma^2 \ll \lambda_2 L_2, \lambda_2 L_1, \lambda_1 L_2$, we get 
a simplified double-slit interference formula,
\begin{equation}
w_2 \approx {\lambda_2(L_1+L_2)\over d} + {\lambda_1L_2\over d}.
\label{young}
\end{equation}
For $\lambda_1=\lambda_2$ we recover the familiar Young's double-slit
interference formula $w_2 = {\lambda_2 D\over d}$, which was obtained for the original ghost interference
experiment \cite{pravatq}, with $D=L_1+2L_2$ being the strange distance between
the double-slit and D2.
For $\lambda_1 \neq \lambda_2$, (\ref{young}) represents the fringe-width
of the ghost interference for photon 2 which strangely also depends on
the wavelength of photon 1.

\subsection{Understanding ghost interference}

Although entanglement between the two photons leads to a virtual double-slit
formation for photon 2, that itself is not enough to yield ghost interference,
just as a real double-slit for photon 1 does not yield an interference.
By virtue of entanglement the two photons are spatially correlated - each one
carries a which-way information about the other. By detecting photons 1
by a {\em fixed} D1, in the region where the wave-packets from the two slits
overlap, one erases the which-way information. Once the which-way information
has been erased, interference can occur \cite{eraser1,eraser2}. Thus, the two-color
ghost interference
is seen only when the photons 2 are detected in coincidence with a fixed D1.

Although the virtual double-slit for particle 2 comes into being only after
particle 2 travels a distance $L_2$ from the source, the particle carries
with itself the phase information of its evolution from the source for the
time $t_0$. Because of coincident counting, the change in phase because of
the evolution of particle 1 is added to that of particle 2. This is what
leads to the unusual fringe-width given by (\ref{young}).

\subsection{Effect of converging lens}

Since Ding et al. \cite{twocolor} have used a converging lens before the
detector D2, the fringe-width formula given by (\ref{young}) doesn't directly
apply. However, if this experiment is done without the converging lens,
the formula (\ref{young}) can be experimentally tested.

\begin{figure}[h]
\centerline{\resizebox{9.0cm}{!}{\includegraphics{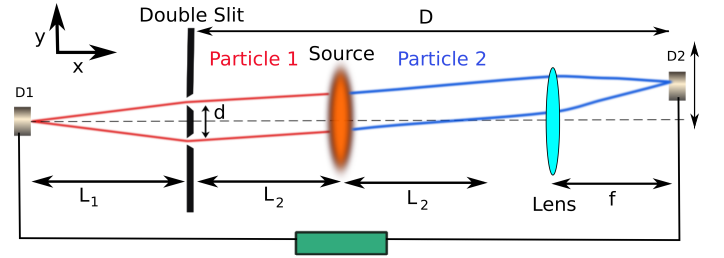}}}
\caption{ The setup for a two-color ghost interference experiment with
a converging lens added. The lens is kept at a distance $f$ before the
detector D2, where $f$ is its focal length. The distance between the
source and the lens is $L_1+L_2-f$.}
\label{lensfig}
\end{figure}

It may be worthwhile to incorporate the effect of a coverging lens in our
theoretical analysis, and see how the results are modified. This will allow
us to make contact with Ding et al.'s experimental results.  In order to
do that, we consider the experimental setup similar to that of
Ding et al. \cite{twocolor} (see Fig. \ref{lensfig}). 
A converging lens of focal length $f$ is kept at distance $f$ before 
detector D2. Consequently, the distance between the source and the lens
is $L_1+L_2-f$.

In our preceding calculation, particles 1 and 2 travel a distance $L_2$
so that particle 1 passes through the double-slit. Immediately after this
the two-particle state is given by eqn. (\ref{virtual}). The situation in
this case is exactly the same, hence (\ref{virtual}) still holds. However,
after this, instead of travelling
a distance $L_1$ to reach D2, particle 2 now travels a distance $L_1-f$
to reach the lens. The two-particle state at this time can be easily
computed to yield
\begin{eqnarray}
\Psi(y_1,y_2,t) &=& 
C_L e^{{{-(y_1-y_0)^2\over\epsilon^2+i\Lambda_1(L_1-f)}}}
 e^{{{-(y_2 - y_0')^2 \over \Gamma+i\Lambda_2(L_1-f)}}}
\nonumber\\
&&+ C_L e^{{{-(y_1+y_0)^2\over\epsilon^2+i\Lambda_1(L_1-f)}}}
 e^{{{-(y_2 + y_0')^2 \over \Gamma+i\Lambda_2(L_1-f)}}},
\label{beforelens}
\end{eqnarray}
where
\begin{equation}
C_L = {1\over \sqrt{\pi}\sqrt{\epsilon+i\Lambda_1L_1\epsilon}
\sqrt{\sqrt{\Gamma_r}+(\Gamma_i+i\Lambda_2L_1)/\sqrt{\Gamma_r}}}.
\end{equation}
and $\Gamma \approx \gamma^2 + i(\Lambda_1+\Lambda_2)L_2$.
Here we have used rescaled wavelengths $\Lambda_1=\lambda_1/\pi,
~\Lambda_2=\lambda_2/\pi$.

The effect of a convegring lens on a general Gaussian wave-packet is 
such that in its subsequent dynamics, it narrows instead of spreading.
If a Gaussian wave-packet of width $\sigma$ starts from a distance
$2f$ from the lens, it should
come back to its original width after a distance $2f$ after
the lens. Also, the observed width of the wavepacket, immediately after
emerging from the lens should be the same as that just before entering
the lens. In general, we can quantify the effect of the lens by a unitary
transformation of the form \cite{popper}
\begin{eqnarray}
  \mathbf{U}_f {(\pi/2)^{-1/4}\over\sqrt{\sigma+{i\Lambda L\over\sigma}}} 
\exp\left({-y_1^2 \over \sigma^2+i\Lambda L}\right) &=& 
{(\pi/2)^{-1/4}\over\sqrt{\tilde{\sigma}+{i\Lambda (L-4f)\over
\tilde{\sigma}}}}\nonumber\\
&&\exp\left({-y_1^2 \over \tilde{\sigma}^2+i\Lambda (L-4f)}\right),
\label{lens} \nonumber\\
\end{eqnarray}
where $L$ is the distance the wave-packet, of an initial width $\sigma$, 
traveled before passing through the lens, and $\tilde{\sigma}$ is such that
it satisfies
\begin{equation}
\tilde{\sigma}^2+{\Lambda^2(L-4f)^2\over\tilde{\sigma}^2} =
\sigma^2+{\Lambda^2 L^2\over\sigma^2}.
\label{sigma}
\end{equation}

We make this transformation on the state (\ref{beforelens}) and let it 
evolve such that particle 2 travels a distance $f$ to reach D2.
The probability density of finding particle 1 at $y_1$ and particle 2 at $y_2$
is given by
%\begin{widetext}
\begin{eqnarray}
P(y_1,y_2) &=& |C_f|^2 \times\nonumber\\ 
&&\left(e^{-{2(y_1-y_0)^2\over\Delta_1} -{2(y_2 - y_0)^2 \over \Delta_2}}
+ e^{-{2(y_1+y_0)^2\over\Delta_1} -{2(y_2 + y_0)^2 \over \Delta_2}}\right.
\nonumber\\
&&\left.+ 2e^{-{2(y_1^2+y_0^2)\over\Delta_1}
-{2(y_2^2 + y_0^2) \over \Delta_2}}
\cos\left[\theta_1 y_1 + \theta_2 y_2\right]\right),
\label{plot2}
\end{eqnarray}
%\end{widetext}
where
$\Delta_1=\epsilon^2+{\lambda_1^2 L_1^2\over\pi^2\epsilon^2}$,
$\Delta_2=\gamma^2+{(\lambda_2 (L-4f)+\lambda_1 L_2)^2\over\pi^2\gamma^2}$,

$\theta_1 = {2d\lambda_1 L_1/\pi\over \epsilon^4 + \lambda_1^2L_1^2/\pi^2}$,
$\theta_2 = {2\pi d[\lambda_2(L-4f)+\lambda_1L_2)]\over \gamma^4\pi^2 +
(\lambda_2(L-4f)+\lambda_1L_2)^2}$,\\
and
$C_f = {1\over \sqrt{\pi}\sqrt{\epsilon+{i\lambda_1L_1\over\pi\epsilon}}
\sqrt{\gamma+{i\lambda_2 (L-4f)+i\lambda_1L_2\over\pi\gamma}}}.$
While arriving at this result, we have ignored the change in $\gamma$ which
was expected from (\ref{sigma}), because it does not affect the quantities
of our interest.

For $\gamma^2 \ll \Lambda_2 L_2, \Lambda_2 L_1, \Lambda_1 L_2$, we get 
a simplified double-slit interference formula,
\begin{equation}
w_2 \approx {\lambda_2(L_1+L_2-4f)\over d} + {\lambda_1L_2\over d}.
\label{young2}
\end{equation}
Eqn. (\ref{young2}) shows that the fringe width will be reduced after
inserting a converging lens.  Ding et al.'s experimental results can be
analyzed using this formula. However, Ding et al. have not mentioned the values
of $L_1, L_2$ etc, but have carried out a simple geometrical analysis.
There is reasonable numerical agreement between Ding et al.'s simple
geometrical analysis and their experimental results. However, we believe
that their experimental results analyzed using eqn. (\ref{young2}) will
give a better agreement.

We also emphasize that formulas (\ref{young}, \ref{young2}) are a signature
of non-classical correlations between photons. An experimental verification of
them would corroborate
the non-classical origin of the two-color ghost interference. For this,
the two-color ghost interference experiment should preferably be carried out
without the converging lens. In addition, the detector D1 should be as narrow
as D2, to prominently bring out the correlation between photons. In
Ding et al.'s experiment, D1 is a bucket detector with width 1 mm, whereas
D2 is a point detector with width 0.2 mm \cite{twocolor}. From the point
of view of our calculation, having a bucket
detector at D1 would amount to averaging over a range of values of $y_1$
in (\ref{plot2}) or (\ref{pattern}).

\section{Conclusion}

To summarize, we have theoretically analyzed the two-color ghost interference
experiment, in a slightly more general setting. We find that the fringe width
of the interference pattern for photon 2, also depends on the wavelength of 
photon 1. This is a completely non-classical feature. With a slight modification
of the experiment, this conclusion can be experimentally tested. Its 
confirmation would be an evidence of the non-classical
origin of the two-color ghost interference.

\begin{acknowledgement}
This work was carried out during a summer project at the Centre for
Theoretical Physics, Jamia Millia Islamia.  Sheeba Shafaq thanks the
Centre for providing her the facilities of the centre during the course of the
summer project.
\end{acknowledgement}

\end{document}